\documentclass[twocolumn,english]{revtex4}
\usepackage{color}
\usepackage{bm}
\usepackage{amsthm}
\usepackage{amsmath}
\usepackage{esint}
\usepackage{graphicx,epsf}
\usepackage{dcolumn}
\usepackage{amsfonts}
\usepackage{mathrsfs}
\usepackage{subfig}

\newcommand{\cb}{\color{blue}}
\newcommand{\crr}{\color{red}}

\begin{document}

\title{Nonlinear coupling of reversed shear Alfv\'en eigenmode
and toroidal Alfv\'en eigenmode during current ramp}

\author{Shizhao Wei$^1$, Yahui Wang$^1$, Peiwan Shi$^{2,3}$,  Wei Chen$^2$, Ningfei Chen$^1$ and Zhiyong Qiu$^1$\footnote{E-mail: zqiu@zju.edu.cn}}
\affiliation{$^1$Institute for Fusion Theory and Simulation and Department of Physics, Zhejiang University, Hangzhou 310027, P.R.C\\
$^2$Southwestern Institute of Physics - P.O. Box 432 Chengdu 610041, P.R.C.\\
$^3$\mbox{Key Laboratory of Materials Modification by Laser, Ion, and Electron Beams (Ministry of Education),} School of Physics, Dalian University of Technology, Dalian 116024, P.R.C.}
\begin{abstract}
Two novel nonlinear mode coupling processes for reversed shear Alfv\'en eigenmode (RSAE) nonlinear saturation are proposed and investigated. In the first process, RSAE nonlinearly couples to a co-propagating toroidal Alfv\'en eigenmode (TAE) with the same toroidal and poloidal mode numbers, and generates a geodesic acoustic mode (GAM). In the second process, RSAE couples to a counter-propagating TAE and generates an ion acoustic wave quasi-mode (IAW). The condition for the two processes to occur is favored during current ramp.  Both processes contribute to effectively saturate the Alfv\'enic instabilities, as well as nonlinearly transfer of energy from energetic fusion alpha particles to fuel ions in burning plasmas.
\end{abstract}
\maketitle

In next generation magnetically confined fusion devices
such as ITER \cite{KTomabechiNF1991}, energetic particles (EPs), e.g., fusion alpha particles, are
expected to play important roles as they contribute significantly to the total power
density and drive symmetry breaking collective modes including shear Alfv\'en wave (SAW) instabilities \cite{AFasoliNF2007, LChenRMP2016,WChenCPL2020}.
SAW can lead to EP anomalous transport loss, degradation of burning plasma
performance and possible damage of plasma facing components \cite{RDingNF2015}. So for
understanding of EP confinement and thus fusion plasma
performance, the in-depth research of SAW instability dynamics including nonlinear evolution is needed.
Due to plasma nonuniformities and equilibrium magnetic field geometry,
SAW instabilities can be  excited as EP modes (EPMs) in the continuum \cite{LChenPoP1994},
or various discrete Alfv\'en eigenmodes (AEs) inside forbidden gaps of SAW continuum
such as toroidal AE (TAE) \cite{CZChengAP1985,LChenVarenna1988,GFuPoFB1989}, reversed shear AE
(RSAE) \cite{HBerkPRL2001,FZoncaPoP2002,GKramerNF2004}, beta-induced AE (BAE) \cite{WHeidbrinkPRL1993}, etc.
The nonlinear mode coupling of SAW instabilities, providing an effective channel for the mode saturation,
has been observed in experiments \cite{PBurattiNF2005}, and investigated analytically \cite{TSHahmPRL1995, FZoncaPRL1995,YTodoNF2010, ABiancalaniPRL2010, LChenPRL2012}
as well as in large scale simulations \cite{YTodoNF2010,HZhangPST2013,ABiancalaniIAEAFEC2018}.
In this work, two possible channels for RSAE saturation as the result  of the nonlinear coupling to TAE during  current ramp are proposed and investigated, which are of relevance for burning plasmas in future reactors due to their typically reversed shear advanced scenarios where RSAEs can be strongly driven unstable by core localized fusion alpha particles \cite{KTomabechiNF1991,XGongNF2019,JHuangNF2020,TWangPoP2018,ZRenNF2020}.

TAE is one of the most well-known SAW eigenmodes and widely exists in present-day and future magnetically confined plasmas.
It can be resonantly excited by EPs in the SAW continuum gap
induced by torodicity \cite{CZChengAP1985,GFuPoFB1989}, as the result of the coupling of neighbouring
poloidal harmonics of SAW continuum, and is typically localized near the center of
two mode rational surfaces, where $q\left(r\right)\simeq\left(2m+1\right)/\left(2n\right)$,
with the parallel wavenumbers of the two dominant poloidal harmonics
both being $\left|k_{\parallel}\right|\simeq1/\left(2qR_{0}\right)$.
Here, $m/n$ are the poloidal/toroidal mode number and $q$ is the
safety factor of the torus. Furthermore, TAE is considered as an important player
in transporting EPs due to its low excitation threshold and suitable
resonance condition with fusion alpha particles with $v_{\alpha}\gtrsim v_{A}$
in ITER \cite{AFasoliNF2007}. Here, $v_{\alpha}$ is the birth velocity of the fusion alpha
particle, and $v_{A}$ is the Alfv\'en speed.

RSAE, also called Alfv\'en cascade in literatures, is excited near
the minima of the safety factor ($q_{min}$), and is often composed by one
dominant poloidal harmonic \cite{HBerkPRL2001,FZoncaPoP2002}. With the parallel wave number $k_{\parallel}\simeq\left|n-m/q_{min}\right|/R$ and thus
its frequency $\omega\simeq k_{\parallel}v_{A}$ determined by the value of
$q_{min}$, RSAE related physics is thus, sensitive to the $q$-profile. One thus expects that, with the change of $q$-profile during, e.g., current
ramp, the RSAE frequency can sweep from BAE (as $\left|nq_{min}-m\right|\simeq0$)
to TAE (as $\left|nq_{min}-m\right|\simeq1/2$) frequency range, which has been shown
in many experiments \cite{MVanZeelandPoP2007} as well as simulations \cite{TWangNF2020}. Furthermore, it is shown in Ref. \cite{MVanZeelandPoP2007} that, as the
core-localized RSAE frequency sweeps up, RSAE can temporally
couple to TAE and generate a global mode, creating an effective channel for particle global
transport from Tokamak core to edge. This channel was also predicted and investigated theoretically in Ref. \cite{FZoncaPoP2002}. The coupling of RSAE and TAE with different mode numbers and generating
a low frequency mode during the RSAE frequency sweeping process has also been observed
in HL-2A experiments \cite{PShiPC2019}.

In this work, we show that, as the RSAE frequency sweeps up during current ramp,
two important potential nonlinear mode coupling processes may happen. In the first process, a RSAE $\Omega_R\equiv\Omega_{R}\left(\omega_{R}, \mathbf{k_{R}}\right)$
couples to a TAE $\Omega_T\equiv\Omega_{T}\left(\omega_{T}, \mathbf{k_{T}}\right)$
with same poloidal and toroidal mode numbers and generates a geodesic acoustic mode (GAM) $\Omega_G\equiv\Omega_{G}\left(\omega_{G}, \mathbf{k_{G}}\right)$ with toroidally symmetric and poloidally near symmetric mode structure \cite{NWinsorPoF1968,FZoncaEPL2008}.
Here, subscripts ``R", ``T" and ``G" represent RSAE, TAE and GAM, respectively. In the second channel, a RSAE couples to a counter-propagating TAE and
generates an ion acoustic wave (IAW) $\Omega_S\equiv\Omega_{S}\left(\omega_{S}, \mathbf{k_{S}}\right)$,
and this process can occur even if RSAE has different poloidal/toroidal mode numbers with TAE.
Therefore, generally speaking, the condition for the latter process to occur, including radial mode structure overlapping, is more easily satisfied,  as we show later. It is worthwhile noting that, in both channels, the low frequency secondary modes, i.e., GAM and IAW, can both be Landau damped due to resonance with thermal ions, and thus heat fuel deuterium and tritium ions, proving a new alpha channelling mechanism that indirectly transfer fusion alpha particle energy to fuel ions \cite{NFischPRL1992,TSHahmPST2015,ZQiuPRL2018}.

We use the standard nonlinear perturbation theory to study
the nonlinear interactions between RSAE and TAE during current ramp.
For typical low-$\beta$ discharges, magnetic compression can be neglected, and the electrostatic potential $\delta\phi$ and the parallel component of vector
potential $\delta A_{\parallel}$ are used as perturbed field variables. Here, $\beta$ is the ratio of thermal to magnetic pressure. Furthermore,
we use $\delta\psi\equiv\omega\delta A_{\parallel}/\left(ck_{\parallel}\right)$
as an alternative field variable replacing $\delta A_{\parallel}$
for convenience, and one can have $\delta\phi=\delta\psi$ in the ideal MHD limit.

For RSAE and TAE with $nq\gg1$ in reactor relevant parameter regime \cite{AFasoliNF2007,LChenRMP2016,TWangPoP2018,ZRenNF2020}, we adopt the following ballooning
mode representation in the $\left(r, \theta, \varphi\right)$ field-aligned
flux coordinates \cite{JConnorPRL1978}
\begin{eqnarray}
\delta\phi & = & Ae^{i\left(n\phi-\hat{m}\theta-\omega t\right)}\underset{j}{\sum}e^{-ij\theta}\Phi\left(x-j\right)+c.c.{\crr,}{\cb.}\nonumber
\end{eqnarray}
Here, $m=\hat{m}+j$ with $\hat{m}$ being the reference poloidal number, $x\equiv nq-m$,
$\Phi$ is the parallel mode structure with the typical radial extension comparable to distance between neighboring mode rational surfaces, and $A$ is the mode envelope amplitude.
Furthermore, $\Omega_{G}$ and $\Omega_{S}$ can be shown as
\begin{eqnarray}
\delta\phi_{G} & = & A_{G}e^{i\left(\int k_{G}dr-\omega_{G}t\right)}+c.c., \nonumber\\
\delta\phi_{S} & = & A_{S}e^{i\left(n_{S}\phi-m_{S}\theta-\omega_{S}t\right)}\Phi_{S} + c.c.. \nonumber
\end{eqnarray}
The IAW is a secondary mode and its parallel mode structure $\Phi_{S}$ is determined by $\Phi_{R}$ and $\Phi_{T}$ \cite{ZQiuNF2016}.
Different from previous works on spontaneously decay of a pump wave into two sidebands, e.g., as in the case of zonal flow excited by SAW \cite{LChenPRL2012,ZQiuNF2017}, the RSAE and TAE investigated here are simultaneously driven unstable by EPs \cite{TWangPoP2018},  and thus, the radial mode structure of
RSAE and TAE may not overlap completely.

The nonlinear interaction of $\Omega_{R}$ and $\Omega_{T}$ can be described by quasi-neutrality condition
\begin{eqnarray}
\dfrac{n_{0}e^{2}}{T_{i}}\left(1+\dfrac{T_{i}}{T_{e}}\right)\delta\phi_{k} & = & \underset{s}{\sum}\left\langle qJ_{k}\delta H_{k}\right\rangle _{s},\label{eq:Q.N.}
\end{eqnarray}
and nonlinear gyrokinetic vorticity equation derived from parallel Amp\'ere's law \cite{LChenRMP2016}
\begin{eqnarray}
&&\notag\dfrac{c^{2}}{4\pi\omega_{k}^{2}}B\dfrac{\partial}{\partial l}\dfrac{k_{\perp}^{2}}{B}\dfrac{\partial}{\partial l}\delta\psi_k + \dfrac{e^{2}}{T_{i}}\left\langle \left(1-J_{k}^{2}\right)F_{0}\right\rangle \delta\phi_{k}
\\&&\notag-\underset{s}{\sum}\left\langle \dfrac{q}{\omega_{k}}J_{k}\omega_{d}\delta H_{k}\right\rangle
\\&&\notag=-i\dfrac{c}{B_{0}\omega_{k}}\underset{\mathbf{k}=\mathbf{k'}+\mathbf{k''}}{\sum}\mathbf{\hat{b}}\cdot\mathbf{k''}\times\mathbf{k'}\left[\dfrac{c^{2}}{4\pi}k_{\perp}''^{2}\dfrac{\partial_{l}\delta\psi_{k'}\partial_{l}\delta\psi_{k''}}{\omega_{k'}\omega_{k''}}\right.
\\&&\left.+\left\langle e\left(J_{k}J_{k'}-J_{k''}\right)\delta L_{k'}\delta H_{k''}\right\rangle \right].\label{eq:vorticity equation}
\end{eqnarray}
Here, $J_{k}\equiv J_{0}\left(k_{\perp}\rho\right)$ with
$J_{0}$ being the Bessel function of zero index, $\rho=v_{\perp}/\Omega_{c}$ is the Larmor radius with $\Omega_{c}$ being the cyclotron frequency, $F_{0}$ is the
equilibrium particle distribution function, $\omega_{d}=\left(v_{\perp}^{2}+2v_{\parallel}^{2}\right)/\left(2\Omega_{c}R_{0}\right)\left(k_{r}\sin\theta+k_{\theta}\cos\theta\right)$
is the magnetic drift frequency, $l$ is the length along the equilibrium
magnetic field line, $\left\langle \cdots\right\rangle $ means velocity
space integration, $\sum_{s}$ is the summation of different
particle species with $s=i, e$ representing ion and electron, and $\delta L_{k}\equiv\delta\phi_{k}-k_{\parallel}v_{\parallel}\delta\psi_{k}/\omega_{k}$.
The three terms on the left hand side of equation (\ref{eq:vorticity equation}) are respectively the field line bending, inertial and curvature coupling terms, dominating the linear SAW physics.
The two terms on the right hand side of equation (\ref{eq:vorticity equation})
correspond to Maxwell (MX) and Reynolds stresses (RS) that contribute to nonlinear mode couplings as
MX and RS doesn't cancel each other \cite{LChenPoP2013}, with their contribution dominating in the radially fast varying inertial layer, and $\sum_{\mathbf{k}=\mathbf{k'}+\mathbf{k''}}$ indicates
the wavenumber and frequency matching condition required for nonlinear mode coupling. $\delta H_k$ is the nonadiabatic particle
response, which can be derived from nonlinear gyrokinetic equation \cite{EFriemanPoF1982}:
\begin{eqnarray}
&&\notag\left(-i\omega_{k}+v_{\parallel}\partial_{l}+i\omega_{d}\right)\delta H_{k} = -i\omega_{k}\dfrac{q}{T}F_{0}J_{k}\delta L_{k}
\\&&-\dfrac{c}{B_{0}}\underset{\mathbf{k}=\mathbf{k'}+\mathbf{k''}}{\sum}\mathbf{\hat{b}}\cdot\mathbf{k''}\times\mathbf{k'}J_{k'}\delta L_{k'}\delta H_{k''}.\label{eq:gyrokinetic equation}
\end{eqnarray}
Here, with the diamagnetic drift related term neglected in the first term on the right hand side of equation (\ref{eq:gyrokinetic equation}), we assume that thermal plasma - dominating the nonlinear coupling process - contribution to RSAE/TAE linear destabilization is negligible, and RSAE/TAE are excited by EPs. Inclusion of thermal ion diamagnetic drift effect is straightforward, and will not change the main physics picture here.
For TAE/RSAE with $\left|k_{\parallel}v_{e}\right|\gg\left|\omega_k\right|\gg\left|k_{\parallel}v_{i}\right|,\:\left|\omega_{d}\right|$,
the linear ion/electron responses can be derived to the leading order as
$\delta H^L_{k, i}=eF_{0}J_k\delta\phi_ k/T_{i}$ and $\delta H^L_{k,e}=-eF_{0}\delta\psi_k/T_{e}$.
Furthermore, one can have, to the leading order, ideal MHD constraint is satisfied, i.e., $\delta\phi_{T}=\delta\psi_{T}$, $\delta\phi_{R}=\delta\psi_{R}$,
by substituting these ion/electron responses of TAE and RSAE into quasi-neutrality condition.

\begin{figure}
\centering
\includegraphics[scale=0.40]{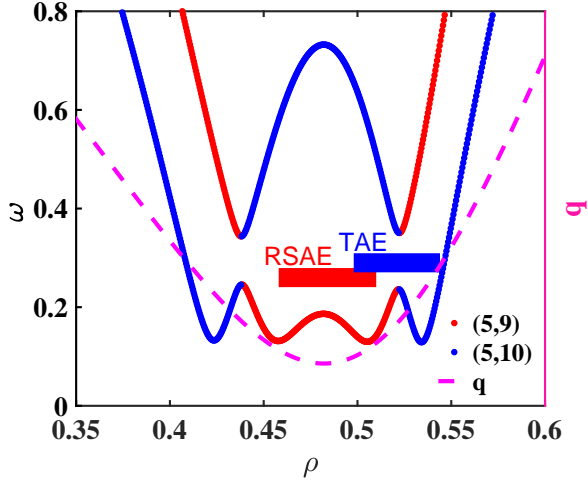}
\caption{Cartoon of GAM generation by RSAE and TAE. The horizontal axis is the normalized minor radius, the dashed purple curve is the $q$-profile, while the red and blue curves are the $m/n=9/5$ and $m/n=10/5$ continua, respectively. The reversed shear $q$-profile has a minimum value at $r\simeq 0.47$, where RSAE is generated above the local maximum of SAW continuum induced by local $q_{min}$. TAE can be generated inside a   radially nearby continuum gap, with the frequency slightly higher than that of the RSAE. These two modes can couple as they are radially overlapped, and generate a low frequency GAM.}
\label{fig:GAM_generation}
\end{figure}

The nonlinear coupling and generation of GAM as the RSAE couples to the TAE can be illustrated in Fig. \ref{fig:GAM_generation}, where a general $n=5$ SAW continuum in a typical reversed shear configuration is given \cite{JBaoJFE2020}. The $m=9$ and $m=10$ continuum are marked in red and blue, respectively, and they coupled at the vicinity of  $q=1.9$. A RSAE can be generated at the local maximum of the continuum, while a TAE can be excited within the continuum gap, with their radial localization determined by $q_{min}$ and $q=(2m+1)/(2n)$, respectively. As $|nq_{min}-m|$ approaches $1/2$ during current ramp, the localization of the RSAE gets closer to that of the TAE, and the frequency difference of RSAE and TAE decreases. As a result, RSAE and TAE may couple, and generate a GAM.  Noting that GAM is characterized by the toroidal/poloidal mode numbers being $n/m=0/0$ and a frequency much lower than those of RSAE/TAE, i.e., $|\omega_G|\ll|\omega_R|,|\omega_T|$, one can have that
TAE and RSAE have opposite poloidal and toroidal mode numbers as well as opposite
frequencies $\left(\omega_{R}\omega_{T}<0\right)$. Thus, the two modes propagate in the same direction, i.e., their parallel phase velocities $V_P\equiv\omega/k_{\parallel}$ have the same sign. For GAM with predominantly electrostatic perturbation, its nonlinear generation can be determined by the vorticity equation, while the quasi-neutrality condition is used in determining the linear polarization of GAM and both AEs \cite{ZQiuPPCF2009,CZChengAP1985}. The linear ion/electron responses
to GAM can be derived considering the $\left|\omega_{tr,e}\right|\gg\left|\omega_{G}\right|\gg\left|\omega_{d,e}\right|$
and $\left|\omega_{G}\right|\gg\left|\omega_{tr,i}\right|, \left|\omega_{d,i}\right|$ orderings, respectively,
and one derives, to the leading order, $\delta H_{G,i}^{L} = eF_{0}J_{G}\delta\phi_{G}/T_{i}$ and $\delta H_{G,e}^{L} = -eF_{0}\overline{\delta\phi_{G}}/T_{e}$.
Here, $\omega_{tr}\equiv v_{\parallel}/(qR_0)$ is the transit frequency, and $\overline{(\cdots)}\equiv \int^{2\pi}_0(\cdots) d\theta/(2\pi)$ denotes flux surface average. The nonlinear equation describing GAM generation can then be derived from nonlinear vorticity equation as
\begin{eqnarray}
&&\notag\dfrac{e^{2}}{T_{i}}\left\langle \left(1-J_{G}^{2}\right)F_{0}\right\rangle \delta\phi_{G}-\underset{s}{\sum}\left\langle \dfrac{q}{\omega}J_{G}\omega_{d}\delta H^L_{G}\right\rangle
\\&&\notag\simeq-i\dfrac{c}{B_{0}\omega}\mathbf{\hat{b}}\cdot\mathbf{k_{R}}\times\mathbf{k_{T}}
\\&&\notag\times\left[\dfrac{c^{2}}{4\pi}\left(k_{T\perp}^{2}-k_{R\perp}^{2}\right)\dfrac{k_{T,\parallel}k_{R,\parallel}}{\omega_{T}\omega_{R}}\delta\psi_{T}\delta\psi_{R}\right.
\\&&\left.+\left\langle e\left(J_{T}-J_{R}\right)\left(\delta\phi_{T}\delta H_{R}+\delta\phi_{R}\delta H_{T}\right)\right\rangle \right].\label{eq:GAM nonlinear vorticity1}
\end{eqnarray}
In deriving equation (\ref{eq:GAM nonlinear vorticity1}),
the linear field line bending term associated with the GAM electromagnetic effects is neglected because GAM is predominantly an electrostatic mode. The last term of equation (\ref{eq:GAM nonlinear vorticity1}) is the RS evaluated using the linear ion responses to RSAE/TAE and the $k^2_{\perp}\rho^2_i\ll1$ ordering.
Noting that $k_{G}=k_{T,r}+k_{R,r}$, $k_{r}\gg k_{\theta}$
for inertial layer responses of TAE and RSAE that dominates the nonlinear mode coupling \cite{LChenPoP2013}, one obtains
\begin{eqnarray}
\notag\delta\phi_{G}&=&\dfrac{c\omega}{B_{0}\left(\omega^{2}-\omega_{G}^{2}\right)}k_{T,\theta}\left(1-\dfrac{k_{T,\parallel}k_{R,\parallel}}{\omega_{T}\omega_{R}}v_{A}^{2}\right)
\\&\times&\left(\delta\phi_{R}\partial_{r}\delta\phi_{T}-\delta\phi_{T}\partial_{r}\delta\phi_{R}\right).\label{eq:GAM nonlinear vorticity2}
\end{eqnarray}
Here, $\omega_{G}$ is the eigenfrequency of GAM and is given as $\omega^2_{G}=\left(7/4+T_{e}/T_{i}\right)v_{ti}^{2}/R_{0}^{2}$ \cite{ZQiuPPCF2009},
$\omega=\omega_{T}+\omega_{R}$ from balancing the temporal evolution of both sides of equation (\ref{eq:GAM nonlinear vorticity1}), which is not necessarily exactly the same as $\omega_{G}$.
Equation (\ref{eq:GAM nonlinear vorticity2}) describes the nonlinear drive of GAM by the beating of co-existing  TAE and RSAE simultaneously driven unstable by e.g., EPs. The effective coupling and GAM generation, requires that $1-k_{T,\parallel}k_{R,\parallel}v^2_A/(\omega_T\omega_R)\neq 0$, i.e., breaking of pure Alfv\'enic state \cite{LChenPoP2013}. This condition is satisfied by the deviation of both RSAE and TAE from ideal SAW dispersion relation, due to the effects of reversed shear profile and toroidicity \cite{FZoncaPoP2002,FZoncaPoP1996}, respectively, and one has $1-k_{T,\parallel}k_{R,\parallel}v^2_A/(\omega_T\omega_R)\simeq O(\epsilon)$ with $\epsilon\equiv r/R_0$ being the inverse aspect ratio. Noting that,
this forced driven process is thresholdless, but only when $\omega_T+\omega_R$ gets sufficiently close to $\omega_{G}$ during RSAE frequency sweeping as a result of current ramp,
GAM will be strongly excited, and the process can be observed experimentally and play an important role in the RSAE/TAE saturation.

In Ref. \cite{MVanZeelandPoP2007}, a central localized RSAE coupling to TAE and generate a global TAE is investigated, where a continuum similar to Fig. \ref{fig:GAM_generation} is used for the illustration of the mechanism. However, the process discussed in Ref. \cite{MVanZeelandPoP2007} is a linear process, where the RSAE has the same toroidal mode number with the TAE while neighboring  poloidal mode number  is coupled through toroidicity, and is thus completely different from the present work.

\begin{figure}[h]
\centering
\includegraphics[scale=0.40]{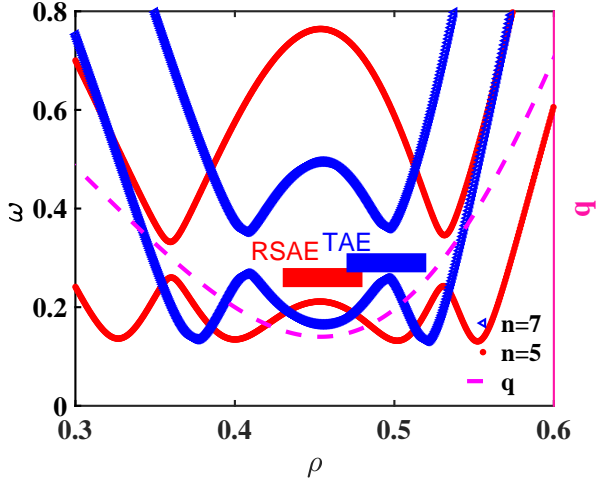}
\caption{Cartoon of IAW generation by RSAE and TAE. The dashed purple curve is the $q$-profile, while the red and blue curves are SAW continua of $n=5$ and $n=7$. An $n=5$ RSAE localizes at the local maximum of the SAW continuum caused by the shear reversal, and an $n=7$ TAE localizes within the SAW continuum induced by toroidicity.}
\label{fig:IAW_generation}
\end{figure}

Considering the other case of RSAE coupling to TAE and
generating an IAW with finite $k_{S,\parallel}$ and $|\omega_{S}|\ll|\omega_{T}|,|\omega_R|$,
one can have TAE and RSAE have opposite frequencies while their parallel wavenumbers have the same sign, and thus are counter-propagating with opposite $V_p$. This process is illustrated in Fig. \ref{fig:IAW_generation} using the nonlinear coupling of an $n=5$ RSAE and an $n=7$ TAE as example, where the red and blue curve represent the $n=5$ and $n=7$ SAW continua,  and the corresponding RSAE and TAE are radially overlapped, leading to potential nonlinear coupling. Note that, the RSAE and TAE can have different toroidal mode numbers, unlike the case of GAM generation limited by the $n_R+n_T=0$ constraint. Thus, it is expected the condition for IAW generation can be more easily satisfied (as shown clearly in Fig. \ref{fig:IAW_generation}), in future burning plasmas where multiple-$n$ SAW instabilities can be driven unstable simultaneously.

Separating the linear to nonlinear particle responses to IAW as
$\delta H_{S}=\delta H_{S}^{L}+\delta H_{S}^{NL}$, the leading order linear ion/electron responses
to IAW can be derived from gyrokinetic equation
as $\delta H_{S,i}^{L} = \left(eF_{0}J_{S}\omega_{S}\delta\phi_{S}\right)/\left[T_{i}\left(\omega_{S}-k_{S,\parallel}v_{\parallel}\right)\right]$ and $\delta H_{S,e}^{L} = 0$, considering the $\left|\omega_{tr,e}\right|\gg\left|\omega_{S}\right|\simeq\left|\omega_{tr,i}\right|\gg\left|\omega_{d,i}\right|, \left|\omega_{d,e}\right|$ ordering.
Furthermore, the nonlinear particle responses to IAW can be derived as
\begin{eqnarray}
\delta H_{S,e}^{NL} & = & -i\dfrac{\hat{\Lambda}}{\omega_{T}}\dfrac{e}{T_{e}}F_{0}\delta\psi_{T}\delta\psi_{R},\label{eq:IAW nonlinear electron}\\
\delta H_{S,i}^{NL} & = & -i\dfrac{\hat{\Lambda}}{\omega_{T}}\dfrac{e}{T_{i}}F_{0}J_{T}J_{R}\delta\psi_{T}\delta\psi_{R},\label{eq:IAW nonlinear ion}
\end{eqnarray}
with $\hat{\Lambda}=\left(c/B_{0}\right)\mathbf{\hat{b}}\cdot\mathbf{k_{R}}\times\mathbf{k_{T}}$. Here, $\omega_S\simeq-\omega_T$ is used.
Assuming ideal MHD condition ($\delta\psi=\delta\phi$) for RSAE and TAE in the radially fast varying inertial layer where nonlinear coupling dominates, the excitation of electrostatic IAW by TAE and RSAE
can be derived from quasi-neutrality condition, by substituting the particle responses into quasi-neutrality condition, and one obtains
\begin{eqnarray}
\mathscr{E}_S\delta\phi_{S} & = & i\dfrac{\hat{\Lambda}}{\omega_{T}}\beta_{S}\delta\phi_{T}\delta\phi_{R}.\label{eq:IAW nonlinear Q.N.}
\end{eqnarray}
Here, $\mathscr{E}_S\equiv1+\tau+\tau\left\langle F_{0}J_{S}^{2}/n_{0}\right\rangle \zeta_{S}Z\left(\zeta_{S}\right)$ is
the linear dispersion function of $\Omega_{S}$, with $\tau\equiv T_{e}/T_{i}$,
$\zeta_{S}\equiv\omega_{S}/\left(k_{S,\parallel}v_{i}\right)$ and
$Z\left(\zeta_{S}\right)$ being the plasma dispersion function defined as
\begin{eqnarray}
Z\left(\zeta_{S}\right) & \equiv\frac{1}{\sqrt{\pi}}\int_{-\infty}^{\infty}\frac{e^{-y^{2}}}{y-\zeta_{S}}dy.\nonumber
\end{eqnarray}
In addition, $\beta_{S}\equiv1+\tau\left\langle F_{0}J_{S}J_{T}J_{R}/n_{0}\right\rangle \left(1+\zeta Z\left(\zeta\right)\right)$ is related to the cross-section of the coupling, and in the case of  counter-propagating  RSAE/TAE of interest here, the RS and MX have the same sign, and   $\beta_S\simeq 2$ in the long wavelength limit.
Similarly with the case discussed before,
equation (\ref{eq:IAW nonlinear Q.N.}) describes the nonlinear generation of IAW by the coupling of TAE and RSAE,
showing that this coupling process is thresholdless,  but only significantly affect the nonlinear dynamics as $|\omega_R+\omega_T|\lesssim O(v_i/(qR_0))$ (i.e., $\zeta_S\lesssim O(1)$). In this parameter regime, RSAE and TAE strongly couples as $|\mathscr{E}_S|\ll 1$, as shown by equation (\ref{eq:IAW nonlinear Q.N.}).  On the other hand, as $\zeta_S\lesssim O(1)$, the nonlinearly generated IAW is a heavily Landau damped quasi-mode that effectively gives energy to thermal ions, leading to effective dissipation of RSAE/TAE wave energy.

In conclusion, two novel nonlinear coupling  channels for RSAE saturation are proposed and investigated. In the first channel, the RSAE couples to a co-propagating TAE with the same toroidal and poloidal mode numbers, and generates a GAM. In the second channel, the RSAE couples to a counter-propagating TAE and generates an IAW. In the latter case, the RSAE and TAE toroidal/poloidal mode numbers are not necessarily the same. The frequency matching condition is more easily satisfied as the RSAE frequency is slightly lower than that of TAE, and is more easily satisfied during  current ramp.
Our results show that, the IAW generation can be more efficient in that 1. the RSAE and TAE don't have to have the same toroidal/poloidal mode numbers, and thus, the condition for mode structure radial overlapping can be more easily satisfied, and 2. the RSAE and TAE are counter-propagating, so the RS and MX will not cancel each other, and thus, the nonlinear coupling cross-section is much larger. In burning plasmas, the RSAE and TAE are linearly excited by EPs simultaneously, and as a result, both processes are thresholdless. In both processes, the nonlinearly generated secondary mode, i.e., GAM and IAW, can  be Landau damped due to resonance with thermal ions, and heat  thermal ions, and thus provides an effectively channel for nonlinearly transfer fusion alpha particles to fuel ions for the sustained burning of fusion plasmas. The saturation level of the low frequency sideband in the two processes, can be estimated from equation(\ref{eq:GAM nonlinear vorticity2}) or (\ref{eq:IAW nonlinear Q.N.}), by expanding the GAM or IAW linear dispersion function along the characteristics \cite{ZQiuPRL2018} and balancing the nonlinear drive by RSAE\&TAE nonlinear coupling, and the GAM/IAW Landau damping. The corresponding fuel ion anomalous heating rate can also be obtained, and will be investigated in a future publication.

This work is supported by the National Key R\&D Program of China  under Grant No. 2017YFE0301900,   the National Science Foundation of China under grant No. 11875233, and China Postdoctoral Science Foundation under grant No. 2020M670756. The authors are grateful to Dr. Jian Bao (Institute of physics, Chinese Academy of Science) for his help with the SAW continua.

%\bibliography{refer}

\end{document}